\documentclass[a4paper, 11pt]{article}

\usepackage{fullpage}

\usepackage{amsfonts}
\usepackage{amssymb}
\usepackage{amsmath}
\usepackage{amsthm}
\usepackage{latexsym}
\usepackage{url}

\newcommand{\norm}[1]{\| #1 \|}

\newcommand{\deltamaps}{\overset{\Delta_1}\longmapsto}
\newcommand{\field}{\mathbb{F}}
\newcommand{\group}{\mathbb{G}}
\newcommand{\Adv}{\mathrm{Adv}^\pm}

\newcommand{\domain}{\mathcal{\widetilde D}}

\newtheorem{theorem}{Theorem}
\newtheorem{lemma}[theorem]{Lemma}
\theoremstyle{remark}
\newtheorem{definition}[theorem]{Definition}
\newtheorem{example}{Example}

\newcommand{\refsec}[1]{Section~\ref{sec:#1}}

\newcommand{\refeqn}[1]{(\ref{eqn:#1})}

\newcommand{\refthm}[1]{Theorem~\ref{thm:#1}}
\newcommand{\reflem}[1]{Lemma~\ref{lem:#1}}

\newcommand{\cD}{{\cal D}}
\newcommand{\tGamma}{{\widetilde{\Gamma}}}

\newcommand{\tG}{{\widetilde{G}}}
\newcommand{\pfstart}{\begin{proof}} 
\newcommand{\pfend}{\end{proof}}

\newcommand{\ignore}[1]{}
\newcommand{\elem}[1]{[\![#1]\!]}

\begin{document}
\title{Adversary Lower Bound for the Orthogonal Array Problem}
\author{Robert \v Spalek%
  \thanks{%
  Google, Inc.}\\
  {\tt spalek@google.com}
}
\date{}
\maketitle

\begin{abstract}
We prove a quantum query lower bound $\Omega(n^{(d+1)/(d+2)})$ for the problem of deciding whether an input string of size $n$ contains a $k$-tuple which belongs to a fixed orthogonal array on $k$ factors of strength $d\le k-1$ and index $1$, provided that the alphabet size is sufficiently large.  Our lower bound is tight when $d=k-1$.

The orthogonal array problem includes the following problems as special cases:
\begin{itemize}
\item $k$-sum problem with $d=k-1$,
\item $k$-distinctness problem with $d=1$,
\item $k$-pattern problem with $d=0$,
\item $(d-1)$-degree problem with $1 \le d \le k-1$,
\item unordered search with $d=0$ and $k=1$, and
\item graph collision with $d=0$ and $k=2$.
\end{itemize}
\end{abstract}

\section{Introduction}

\subsection{History}

One of two main techniques for proving lower bounds on quantum query complexity of Boolean functions is the \emph{adversary method} developed by Ambainis~\cite{ambainis:lowerb, ambainis:degree-vs-qc} and independently by Barnum, Saks, and Szegedy~\cite{bss:semidef} as a generalization of the ``hybrid argument'' introduced by Bennett, Bernstein, Brassard, and Vazirani~\cite{bbbv:hybrid} for the Or function.

The adversary bound was strengthened by H\o yer, Lee, and \v Spalek~\cite{hls:madv} to the negative-weight adversary lower bound, which we define in~\refsec{def}. This stronger version was proved to be optimal by Reichardt~\cite{Reichardt10advtight}, and shown to apply to non-Boolean functions and also to the more general setting of state generation and conversion by Lee, Mittal, Reichardt, \v Spalek, and Szegedy~\cite{lmrss:state-conversion}. Although the negative-weight adversary lower bound is known to be tight, for a long time it had not been used to prove lower bounds for explicit functions. Vast majority of lower bounds by the adversary method used the old positive-weight version of this method, and the only bounds which utilized the power of negative weights were for functions on a constant number of bits and their compositions, and these bounds were obtained by numeric optimization.

The other main technique for quantum query lower bounds is the \emph{polynomial method} developed by Beals, Buhrman, Cleve, Mosca, and de Wolf~\cite{bbcmw:polynomialsj}.  This method is in general incomparable to the adversary method.  Ambainis showed several iterated functions for which the adversary method gives polynomially larger bounds~\cite{ambainis:degree-vs-qc}.  On the other hand, the polynomial method gives stronger bounds for low-error and zero-error algorithms~\cite{bcwz:qerror}.

Another example where the polynomial method used to give stronger bounds than the adversary method is the \emph{element distinctness} function. The input to the function is a string of length $n$ of symbols in an alphabet of size $q$, i.e., $x = (x_1, \dots, x_n)\in [q]^n$. We use notation $[q]$ to denote the set $\{1,\dots,q\}$. The element distinctness function evaluates to 0 if all symbols in the input string are pairwise distinct, and to 1 otherwise.
The quantum query complexity of element distinctness is $O(n^{2/3})$ with the algorithm given by Ambainis~\cite{ambainis:eldist}. Tight lower bounds were given by Aaronson and Shi~\cite{as:collision}, Kutin~\cite{kutin:collision}, and Ambainis~\cite{ambainis:collision} using the polynomial method. 

The positive-weight adversary bound, however, fails for element distinctness. The reason is that this function has 1-certificate complexity 2, and the so-called \emph{certificate complexity barrier}~\cite{ss:adversary, zhang:ambainis} implies that for any function with 1-certificate complexity bounded by a constant, the positive-weight adversary method cannot achieve anything better than $\Omega(\sqrt{n})$.  The negative-weight adversary bound is not limited by this barrier~\cite{hls:madv}, but showing an explicit adversary lower bound breaking the certificate complexity barrier for this function or, in fact, for any function on more than a constant number of bits was open for a long time.

Belovs and \v Spalek~\cite{bs:k-sum} were the first to show such an explicit lower bound.  They re-proved the $\Omega(n^{2/3})$ lower bound for element distinctness using the negative-weight adversary method, and generalized it to the following problem.  
Let $\group$ be a finite Abelian group, $t\in \group$ its arbitrary element, and $k$ an arbitrary but fixed constant. The {\em k-sum problem} consists in deciding whether the input string $x_1,\dots,x_n\in \group$ contains a subset of $k$ elements that sums up to $t$.  This problem was first posed by Childs and Eisenberg~\cite{ce:subset-finding}, who noted that it is the hardest problem among all problems with 1-certificate complexity $k$, because knowledge of any $k-1$ input values doesn't reveal any information about whether that $(k-1)$-tuple can be a part of a 1-certificate or not, and they conjectured that its complexity is $\Omega(n^{k/(k+1)})$.  Belovs and \v Spalek~\cite{bs:k-sum} resolved this conjecture in the positive.

The $\Omega(n^{k/(k+1)})$ lower bound for the $k$-sum problem is tight thanks to the quantum algorithm based on quantum walks on the Johnson graph \cite{ambainis:eldist}. This algorithm was first designed to solve the \emph{$k$-distinctness problem}. This problem asks for detecting whether the input string $x\in [q]^n$ contains $k$ elements that are all equal. Element distinctness is the same as 2-distinctness.  Soon enough it was realized that the same algorithm works for any function with 1-certificate complexity $k$~\cite{ce:subset-finding}, in particular, for the $k$-sum problem. The quantum query complexity of this algorithm is $O(n^{k/(k+1)})$, and the algorithm is thus optimal for the $k$-sum problem.

Quantum walk on the Johnson graph is not optimal for the $k$-distinctness problem when $k>2$.  Belovs and Lee showed a ground-breaking quantum algorithm for a certain promise version of this problem based on learning graphs running in $O(n^{1-2^{k-2}/(2^k-1)}) = o(n^{3/4})$ queries~\cite{bl:k-distinctness}, which Belovs then improved to an algorithm for full $k$-distinctiness~\cite{belovs:learningKDist} with the same query complexity.  However, none of these two algorithms is time-efficient. Very recently, Belovs and independently Childs, Jeffery, Kothari, and Magniez described two new quantum walk algorithms for 3-distinctness, not based on learning graphs, running in time $\tilde O(n^{5/7})$~\cite{belovs:electric-networks, cjkm:3-distinctness}.  The best known lower bound for the $k$-distinctness problem is just $\Omega(n^{2/3})$, by a reduction from element distinctness.

\subsection{Our result}

In this paper, we generalize $k$-distinctness, $k$-sum, and several other problems, and express them as special cases of a general family of functions, characterized by orthogonal arrays.  Let us define orthogonal arrays first.  We use the following notation.  For an $x=(x_1, \dots, x_n)\in X^n$ and $S\subseteq [n]$, let $x_S$ denote the projection of $x$ on $S$, i.e., the vector $(x_{s_1},\dots,x_{s_\ell})$ where $s_1,\dots,s_\ell$ are the elements of $S$ in the increasing order.

\goodbreak
\begin{definition}[Orthogonal array~\cite{rao1947factorial, hss:orthogonal-arrays}]\label{def:orthogonalArray}
Let $X$ be an alphabet.  Assume $T$ is a subset of $X^k$ of size $\lambda \cdot |X|^d$ for integers $0 \le d \le k-1$ and $\lambda \ge 1$.  We say that $T$ is a \emph{$d$-$(X,k,\lambda)$ orthogonal array} iff, for every subset of indices $D \subset [k]$ of size $d$ and for every vector $(y_1, \dots, y_d) \in X^d$, there exist exactly $\lambda$ strings $(x_1, \dots, x_k) \in T$ such that $x_D = y$.  We call $d$ the \emph{strength}, $k$ the \emph{number of factors},  $\lambda$ the \emph{index of the array}, and $|X|$ the \emph{alphabet size}.  We call $T$ \emph{linear} if $X$ is a finite field, and the elements of $T$ form a subspace of the vector space $X^k$.
\end{definition}

In this paper, we restrict ourselves to orthogonal arrays of index $\lambda=1$.

\begin{definition}[Consistent collection of orthogonal arrays]
Assume that each subset $S$ of $[n]$ of size $k$ is equipped with a $d$-$(X, k, 1)$ orthogonal array $T_S$.   A \emph{collection $\{T_S\}_S$ of orthogonal arrays} is called \emph{consistent} iff, for every pair of subsets $S_1, S_2 \subset [n]$ of size $k$ with $|S_1 \cap S_2| \ge d$, their corresponding orthogonal arrays are consistent.  We say that $T_{S_1}$ is consistent with $T_{S_2}$ iff, for every $D \subseteq S_1 \cap S_2$ of size $d$ and every vector $(y_1, \dots, y_d) \in X^d$, the unique vectors $x^1 \in T_{S_1}$ and $x^2 \in T_{S_2}$ satisfying $x^1_D = x^2_D = y$ are consistent on the whole intersection $S_1 \cap S_2$, i.e., $x^1_{S_1 \cap S_2} = x^2_{S_1 \cap S_2}$.
\end{definition}

\begin{definition}[Orthogonal array problem]
Let $\{T_S\}_S$ be a collection of $d$-$(X,k,\lambda)$ orthogonal arrays.
The {\em $d$-$(X,k,\lambda)$ orthogonal array problem} consists in finding an element of any of the orthogonal arrays in the input string. More precisely, the input $x\in X^n$ evaluates to 1 iff there exists a subset $S\subseteq [n]$ of size $k$ such that $x_S \in T_S$.  If the collection is consistent, we call the problem a \emph{consistent orthogonal array problem}.
\end{definition}

The orthogonal array problem was first defined by Belovs and \v Spalek~\cite{bs:k-sum} as a convenient tool to prove a tight lower bound for the $k$-sum problem, and it was also used by Belovs and Rosmanis~\cite{br:power-learning-graphs} to prove a lower bound on the quantum query complexity of certificate structures.  Both these papers only use a special case of orthogonal arrays with strength $k-1$, whereas we allow for any strength $d\le k-1$.

Consider the following orthogonal array problems.  The first three examples have been widely studied before.  The last two examples are new, at least in the context of this paper.

\begin{example}[$k$-distinctness problem~\cite{ambainis:eldist, belovs:learningKDist}]\label{exm:k-distinctness}
Let $X$ be any alphabet.  $T = \{ x^k : x \in X\}$ is a 1-$(X, k, 1)$ orthogonal array.  A collection of these arrays is consistent.
\end{example}

\begin{example}[$k$-sum problem~\cite{ce:subset-finding, bs:k-sum}]\label{exm:k-sum}
Let $\group$ be an Abelian group and $t\in\group$.  $T = \{(x_1, \dots, x_k) \in \group^k: \sum_{i=1}^k x_i = t\}$ is a $(k-1)$-$(\group, k, 1)$ orthogonal array.  A collection of these arrays is consistent.
\end{example}

\begin{example}[Unordered search~\cite{bbbv:hybrid, grover:searchj}]\label{exm:search}
Let $X$ be any alphabet and $x\in X$. $T = \{ x \}$ is a 0-$(X, 1, 1)$ orthogonal array.  Unordered search is equal to the $1$-sum problem.
\end{example}

\begin{example}[$k$-pattern problem]
Let $X$ be any alphabet.  For each $k$-tuple $S$, fix a string $y^S \in X^k$.  $T_S = \{ y^S \}$ is a $0$-$(X, k, 1)$ orthogonal array.

If the collection $\{T_S\}_S$ of the orthogonal arrays is consistent, then the $k$-pattern problem is equivalent to $k$ unordered searches without replacement, because there exists a unique vector $y \in X^n$ such that $y^S = y_S$.  If the collection is inconsistent, then the $k$-pattern problem is more general than unordered search.  For example, the \emph{graph collision} problem~\cite{mss:triangle} is a special case of the 2-pattern problem.  See our open problems for a more detailed discussion.
\end{example}

\begin{example}[$d$-degree problem]\label{exm:d-degree}
Let $\field$ be a finite field and $0 \le d \le k-2$.  For each $k$-tuple $S$, let
$T_S = \{ x_S\in \field^k : \exists \alpha_0, \dots \alpha_d\in\field: \forall s\in S: x_s = \sum_{i=0}^d \alpha_i s^i \}$.  $T_S$ is a linear $(d+1)$-$(\field, k, 1)$ orthogonal array.

A collection of these orthogonal arrays is consistent thanks to the way we consistently use the indices $s\in S$ as the points at which the polynomials are evaluated.  Had we, for example, instead sorted the elements of $S$ in an increasing order, indexed them by $[k]$, and defined the polynomial at these points, we would have obtained a different collection of $(d+1)$-$(\field, k, 1)$ orthogonal arrays, which is not consistent.
\end{example}

$k$-distinctness and $k$-sum represent two extreme examples of orthogonal array problems, differing by their strength, and the $d$-orthogonal array problem naturally interpolates between them.  Given that the quantum query complexity of the $k$-sum problem is known and the complexity of $k$-distinctness is open, it is natural to ask how large lower bound can one prove for the $d$-orthogonal array problem, as a function of $d$.  We address this question and prove the following result.

\begin{theorem}[Main result]
\label{thm:orthogonal}
For a fixed $k$ and $0\le d \le k-1$, an alphabet $X$, and any collection of $d$-$(X,k,1)$ orthogonal arrays $T_S$, the quantum query complexity of the $d$-$(X,k,1)$ orthogonal array problem is $\Omega(n^{(d+1)/(d+2)})$ provided that $|X| \ge n^{k/(k-d)}$. The constant behind the big-Omega depends on $k$ and $d$, but not on $n$, $|X|$, or the choice of $T_S$.  The collection of orthogonal arrays may or may not be consistent.
\end{theorem}

The proofs in our paper are straightforward extensions of the corresponding proofs of the quantum query lower bound for the $k$-sum problem~\cite{bs:k-sum}.


Our lower bound for $k$-distinctness is direct, meaning that it doesn't use reduction from element distinctness, and it gives the same bound $\Omega(n^{2/3})$.  The lower bound for the $d$-orthogonal array problem grows with growing $d$ until it reaches its maximal value $\Omega(n^{k/(k+1)})$ for the $k$-sum problem, where the bound is optimal.  We don't know whether our bound is optimal for any $d < k-1$.

We conjecture that any consistent $d$-$(X, k, 1)$ orthogonal array problem can be solved in $o(n^{k/(k+1)})$ quantum queries when $d < k-1$, using learning graphs like in~\cite{belovs:learningKDist}.  That includes the $(d-1)$-degree problem.  Finding such an algorithm is one of our open problems. 

\section{Adversary Lower Bound}
\label{sec:def}
In this paper we are interested in the quantum query complexity of the $d$-$([q], k, 1)$ orthogonal array problem.  For the definitions and main properties of quantum query complexity refer to, e.g., Ref.~\cite{buhrman&wolf:dectreesurvey}.  For the history, definitions, and relationships between various quantum query lower-bound methods refer to, e.g., Ref.~\cite{hs:survey-lb}.  For the purposes of our paper, it is enough to define the adversary bound, which we do in this section.

We use the formulation from Ref.~\cite{bs:k-sum}.  Compared to the original formulation of the negative-weight adversary bound~\cite{hls:madv}, this formulation is different in two aspects.  First, in order to simplify the notation, we call an adversary matrix a matrix with rows labeled by positive inputs, and columns by negative inputs. It is a quarter of the original adversary matrix that completely specifies the latter. Second, due to technical reasons, we allow several rows to be labeled by the same positive input. All this is captured by the following definition and theorem.

\begin{definition}
\label{defn:adversary}
Let $f$ be a function $f\colon \cD\to \{0,1\}$ with domain $\cD\subseteq [q]^n$.  Let $\domain$ be a set of pairs $(x,a)$ with the property that the first element of each pair belongs to $\cD$, and $\domain_i = \{(x,a)\in \domain : f(x)=i\}$ for $i\in\{0,1\}$.  An {\em adversary matrix} for the function $f$ is a non-zero real $\domain_1\times\domain_0$ matrix $\Gamma$. For an $i\in[n]$, let $\Delta_i$ denote the $\domain_1\times \domain_0$ matrix defined by
\[ \Delta_i\elem{(x,a),(y,b)} = \begin{cases} 0,& x_i=y_i; \\ 1,&\text{otherwise}. \end{cases} \]
\end{definition}

\goodbreak
\begin{theorem}[Adversary bound \cite{hls:madv, bs:k-sum}] \label{thm:adv}
In the notation of Definition~\ref{defn:adversary}, $Q_2(f)=\Omega(\Adv(f))$, where
\begin{equation}\label{eqn:adversary}
\Adv(f) = \max_{\Gamma} \frac{\norm{\Gamma}}{\max_{i\in n} \norm{\Gamma\circ\Delta_i} }
\end{equation}
where the maximization is over all adversary matrices for $f$, $\norm{\cdot}$ is the spectral norm, and $Q_2(f)$ is the quantum query complexity of $f$.
\end{theorem}

\section{Proof}
In this section we prove \refthm{orthogonal} using the adversary lower bound, \refthm{adv}.  The idea of our construction is to embed the adversary matrix $\Gamma$ into a slightly larger matrix $\tGamma$ with additional columns. Then $\Gamma\circ \Delta_i$ is a sub-matrix of $\tGamma\circ \Delta_i$, hence, $\|\Gamma\circ \Delta_i\|\le \|\tGamma\circ \Delta_i\|$.  (In this section we use $\Delta_i$ to denote all matrices defined like in Definition~\ref{defn:adversary}, with the size and the labels of the rows and columns clear from the context.)  It remains to prove that $\|\tGamma\|$ is large, and that $\|\Gamma\|$ is not much smaller than $\|\tGamma\|$.

The proof is organized as follows.  In \refsec{tGamma} we define $\tGamma$ depending on certain parameters $\alpha_m$, in \refsec{normtGamma} we analyze its norm, in Sections~\ref{sec:Delta1} and~\ref{sec:tGamma1Norm} we calculate $\norm{\tGamma\circ\Delta_i}$, in \refsec{alpha} we optimize $\alpha_m$, and, finally, in \refsec{Gamma} we prove that the norm of the true adversary matrix $\Gamma$ is not much smaller than the norm of $\tGamma$.

\subsection{Adversary matrix}
\label{sec:tGamma}
Matrix $\tGamma$ consists of $\binom n k$ matrices $\tG_{s_1,\dots,s_k}$ stacked one on another for all possible choices of subset $S=\{s_1,\dots,s_k\}\subset[n]$:
\begin{equation}\label{eqn:tGamma}
\tGamma = \left(
\begin{array}{c} \tG_{1,2,\dots,k} \\ \tG_{1,2,\dots,k-1,k+1} \\ \dots \\ \tG_{n-k+1,n-k+2,\dots,n} \\ \end{array}
\right) \enspace.
\end{equation}
Each $\tG_S$ is a $q^{n-k+d} \times q^n$ matrix with rows indexed by inputs $(x_1, \dots, x_n) \in [q]^n$ such that $x_S \in T_S$, and columns indexed by all possible inputs $(y_1, \dots, y_n) \in [q]^n$.

We say that a column with index $y$ is {\em invalid} if $y_S\in T_S$ for some $S\subseteq[n]$. After removing all invalid columns, $\tG_S$ will represent the part of $\Gamma$ with the rows indexed by the inputs having an element of the orthogonal array on $S$.  Note that some positive inputs appear more than once in $\Gamma$. More specifically, an input $x$ appears as many times as many elements of the orthogonal arrays it contains. 

This construction may seem faulty, because there are elements of $[q]^n$ that are used as labels of both rows and columns in $\tGamma$, and hence, it is trivial to construct a matrix $\tGamma$ such that the value in~\refeqn{adversary} is arbitrarily large. However, we design $\tGamma$ in a specifically restrictive way so that it still is a good adversary matrix after the invalid columns are removed. 

Let $J_q$ be the $q\times q$ all-ones matrix. Assume $e_0,\dots,e_{q-1}$ is an orthonormal eigenbasis of $J_q$ with $e_0=1/\sqrt{q}\cdot (1,\dots,1)$ being the eigenvalue $q$ eigenvector. Consider the vectors of the following form:
\begin{equation}
\label{eqn:v}
v = e_{v_1}\otimes e_{v_2}\otimes\cdots\otimes e_{v_n},
\end{equation}
where $v_i \in \{0,\dots,q-1\}$. These are eigenvectors of the Hamming Association Scheme on $[q]^n$. For a vector $v$ from~\refeqn{v}, the {\em weight} $|v|$ is defined as the number of non-zero entries in $(v_1,\dots,v_n)$. Let $E_k^{(n)}$, for $k=0,\dots,n$, be the orthogonal projector onto the space spanned by the vectors from~\refeqn{v} having weight $k$. These are the projectors on the eigenspaces of the association scheme. Let us denote $E_i = E^{(1)}_i$ for $i=0,1$. These are $q\times q$ matrices. All entries of $E_0$ are equal to $1/q$, and the entries of $E_1 = I - E_0$ are given by
\[
E_1\elem{x,y} = \begin{cases}
1-1/q,& x=y;\\
-1/q,& x\ne y.
\end{cases}
\]

Elements of $S$ in $\tG_S$ should be treated differently from the remaining elements. For them, we define a $q^d\times q^k$ matrix $F_S$. It has rows labelled by the elements of $T_S$ and columns by the elements of $[q]^k$, and is defined as follows.

\begin{definition}\label{def:F}
Let 
\[E^{(k)}_{\le d} = \sum_{i=0}^d E^{(k)}_i = \sum_{\substack{u = e_{u_1} \otimes \cdots \otimes e_{u_k}\\ |u|\le d}} uu^*\]
 be the projector onto the subspace spanned by the vectors of weight at most $d$.  Let $F_S$ be $q^{(k-d)/2}$ times the sub-matrix of $E^{(k)}_{\le d}$ consisting of only the rows from $T_S$. 
\end{definition}


Finally, we define $\tGamma$ as in~\refeqn{tGamma} with $\tG_S$ defined by
\begin{equation}
\label{eqn:GT}
\tG_S = \sum_{m=0}^{n-k} \alpha_m F_S \otimes E^{(n-k)}_m \enspace,
\end{equation}
where $F_S$ acts on the elements in $S$ and $E_m$ acts on the remaining $n-k$ elements. The coefficients $\alpha_m$ will be specified later.

\subsection{Norm of {\large $\tGamma$}}
\label{sec:normtGamma}

\begin{lemma}\label{lem:normGamma}
Let $\tGamma$ be like in~\refeqn{tGamma} with $\tG_S$ defined as in~\refeqn{GT}. Then
\begin{itemize}
\item[(a)] $\|\tGamma\| = \Omega(\alpha_0 n^{k/2})$,
\item[(b)] $\|\tGamma\| = O(\max_m \alpha_m n^{k/2})$.
\end{itemize}
\end{lemma}

\pfstart
Fix a subset $S$ and denote $T=T_S$ and $F=F_S$. Recall that $E^{(k)}_{\le d}$ is the sum of $uu^*$ over all $u=e_{u_1}\otimes\cdots\otimes e_{u_k}$ with at least $k-d$ elements $u_j$ equal to 0, and $F$ is the restriction of $E^{(k)}_{\le d}$ to the rows in $T$.

For $u=e_{u_1}\otimes\cdots\otimes e_{u_k}$ and $L \subset [k]$ of size $|L|=k-d$ such that $u_L = e_0^{\otimes(k-d)}$, let $u^L$ denote the $q^{(k-d)/2}$ multiple of $u$ restricted to the elements in $T$. The reason for the superscript is that we consider the following process of obtaining $u^L$: we treat $T$ as $[q]^d$ by erasing the elements indexed by $L$ in any string of $T$, then $u^L$ coincides on this set with $u$ with the $L$-terms removed. 

In this notation, the contribution from $uu^*$ to $F$ equals $u^{L_u}u^*$, where $L_u$ is any set of $k-d$ positions in $u$ containing $e_0$. In general, we do not know how the $u^L$ vectors relate for different $L$. However, we know that, for a fixed $L$, they are all orthogonal; and for any $L$, $(e_0^{\otimes k})^L$ is the vector $1/\sqrt{q^d}\cdot (1,\dots,1)$.

Let us start with proving (a). We estimate $\|\tGamma\|$ from below by $w^*\tGamma w'$, where $w$ and $w'$ are unit vectors with all elements equal. In other words, $\|\tGamma\|$ is at least the sum of all its entries divided by $\sqrt{\binom n k q^{2n+d-k}}$. In order to estimate the sum of the entries of $\tGamma$, we rewrite~\refeqn{GT} as
\begin{equation}
\label{eqn:GT2}
\tG_S = \alpha_0 e_0^{\otimes(n+d-k)}(e_0^{\otimes n})^* + {\sum_{u,v}} \alpha_{|v|} (u^{L_u}\otimes v)(u\otimes v)^* \enspace,
\end{equation}
where the summation is over all $u$ and $v$ such that at least one of them contains an element different from $e_0$. The sum of all entries in the first term of~\refeqn{GT2} is $\alpha_0 q^{n+(d-k)/2}$. The sum of each column in each of  $(u^{L_u}\otimes v)(u\otimes v)^*$ is zero because at least one of $u^{L_u}$ or $v$ sums up to zero. By summing over all $\binom n k$ choices of $S$, we get that $\norm{\tGamma} \ge \alpha_0\sqrt{\binom n k} = \Omega(\alpha_0 n^{k/2})$.

In order to prove (b), express $F_S$ as $\sum_{L \subset [k]: |L|=k-d} F^L_S$ with $F^L_S = \sum_{u\in U_L} u^L u^*$. Here $\{U_L\}$ is an arbitrary decomposition of all $u$ such that $U_L$ contains only $u$ with $e_0$ in the $L$-positions. Define $\tG^L_S$ as in~\refeqn{GT} with $F_S$ replaced by $F^L_S$, and $\tGamma^L$ as in~\refeqn{tGamma} with $\tG_S$ replaced by $\tG^L_S$.

Since all $u^L$ vectors are orthogonal for a fixed $L$, we get that
\[
(\tG^L)^*\tG^L = \sum_{u\in U_L, v} \alpha_{|v|}^2 (u\otimes v)(u\otimes v)^*,
\]
thus $\| (\tG^L)^*\tG^L \| = \max_m \alpha_m^2$. By the triangle inequality, 
\[
\|\tGamma^L\|^2 = \left\|(\tGamma^L)^*\tGamma^L\right\| =  \left\|\sum_S (\tG_S^L)^*\tG_S^L \right\| \le 
\binom n k \max_m \alpha_m^2. 
\]
Since $\tGamma = \sum_{L \subset [k]: |L|=k-d} \tGamma^L$ and $\binom k {k-d} = O(1)$, another application of the triangle inequality finishes the proof of (b).
\pfend

\subsection{Action of {\large $\Delta_1$}}
\label{sec:Delta1}

The adversary matrix is symmetric in all input variables and hence it suffices to only consider the entry-wise multiplication by $\Delta_1$.  Precise calculation of $\|\tGamma \circ \Delta_1\|$ is very tedious, but one can get an asymptotically tight bound using the following trick.
Instead of computing $\tGamma \circ \Delta_1$ directly, we arbitrarily map $\tGamma \deltamaps \tGamma_1$ such that $\tGamma_1 \circ \Delta_1 = \tGamma \circ \Delta_1$, and use the inequality $\|\tGamma_1 \circ \Delta_1\| \le 2 \|\tGamma_1\|$ that holds thanks to $\gamma_2(\Delta_1)\le 2$ \cite{lmrss:state-conversion}.  In other words, we change arbitrarily the entries with $x_1 = y_1$.  We use the mapping
\begin{equation}\label{eqn:E1}
E_0 \deltamaps E_0 \enspace, \qquad E_1 \deltamaps -E_0 \enspace.
\end{equation}
The projector $E^{(k)}_{\le d}$ is mapped by $\Delta_1$ as
\begin{eqnarray}
E^{(k)}_{\le d} &\deltamaps& E_0 \otimes E^{(k-1)}_d \enspace. \\
\noalign{\noindent It follows that\medskip}
F &\deltamaps& 
F_1 = \sum_{\substack{u=e_{u_1}\otimes\cdots\otimes e_{u_k}\\ u_1=0, |u|=d}} u^{L_u} u^* \enspace,
\label{eqn:F1} \\
\noalign{\noindent where $u^{L_u}$ is defined like in the proof of \reflem{normGamma}.  For a subset $L\subset [k]$ of size $|L|=k-d$ that contains $1\in L$,\medskip}
F^L &\deltamaps& 
(F^L)_1 = \sum_{\substack{u \in U_L\\ |u|=d}} u^L u^* \enspace,
\label{eqn:FL1}
\end{eqnarray}
where $F^L$ and $U_L$ is defined like in the proof of \reflem{normGamma}(b).

\subsection{Norm of {\large $\tGamma_1$}}
\label{sec:tGamma1Norm}
\begin{lemma}\label{lem:normGamma1}
Let $\tGamma$ be like in~\refeqn{tGamma} with $\tG_S$ defined as in~\refeqn{GT}, and map $\tGamma \deltamaps \tGamma_1$, $\tG_S \deltamaps (\tG_S)_1$, and $F_S \deltamaps (F_S)_1$ using \refeqn{E1} and \refeqn{F1}. Then 
\[\|\tGamma_1\| = O\left(\max_m \left(\max (\alpha_m m^{d/2} n^{(k-1-d)/2},
(\alpha_m - \alpha_{m+1}) n^{k/2})\right)\right).\]
\end{lemma}

\pfstart
\renewcommand{\S}{\mathcal{S}}
In order to prove the upper bound, we express $\tGamma_1 = \sum_L \tGamma^L _1$, $(\tG_S)_1 = \sum_L (\tG_S^L)_1$, and $(F_S)_1 = \sum_{L\subset[k]: |L|=k-d} (F_S^L)_1$, like in the proof of~\reflem{normGamma}(b), and upper-bound each $\|\tGamma_1^L\|$ separately.  Note that even though $L$ is a subset of $[k]$ and not $S$, we can still use $L$ to select a subset of elements of $S$ if each $S$ is ordered in the ascending order. $S_L = \{s_i : i\in L\}$ for $S = \{s_1, \dots, s_k\}$.

We have $\|\tGamma^L_1\|^2 = \|(\tGamma^L_1)^* \tGamma^L_1\| = \|\sum_S (\tG_S^L)_1^* (\tG_S^L)_1\|$.
Decompose the set of all possible $k$-tuples of indices into $\S_1 \cup \S_2$, where $\S_1$ are $k$-tuples containing 1 and $\S_2$ are $k$-tuples that don't contain 1.  We upper-bound the contribution of $\S_1$ to $\|\tGamma_1^L\|^2$ by $\max_m \alpha_m^2 \binom {m+d} d \binom {n-m-d-1} {k-1-d}$ and the contribution of $\S_2$ by $\max_m(\alpha_m - \alpha_{m+1})^2 \binom {n-1} k$, and apply the triangle inequality.

Let $v = e_{v_1} \otimes \cdots \otimes e_{v_n}$ with $|v| = m+d$, and let $S \in \S_1$.  Then, by~\refeqn{FL1},
\[
(\tG_S^L)_1 v = \left\{ \begin{tabular}{l l}
$\alpha_m v^{S_L}$, & $v_1=0$, $|v_S|=d$, and $|v_{S_L}| = 0$ \\
0, & otherwise.
\end{tabular} \right.
\]
Here $v_S = \bigotimes_{s\in S} e_{v_s}$ and $v^{S_L} = q^{(k-d)/2} \bigotimes_{i \in [n] - S_L} e_{v_i}$.

For different $v$, these are orthogonal vectors, and hence $v$ is an eigenvector of $(\tG_S^L)_1^* (\tG_S^L)_1$ of eigenvalue $\alpha_m^2$ if $v_1=0$, $|v_S|=d$, and $|v_{S_L}|=0$, and of eigenvalue 0 otherwise. For every $v$ with $v_1=0$ and $|v|=m+d$, there are $\binom {m+d} d \binom {n-m-d-1} {k-1-d}$ sets $S\in \S_1$ such that $|v_S|=d$, and hence at most as many sets $S\in \S_1$ such that $(\tG_S^L)_1 v \ne 0$. We apply the triangle inequality, and conclude that the contribution of $\S_1$ is as claimed.

Now consider an $S\in \S_2$, that means $1 \not\in S$.
\begin{eqnarray*}
\tG_S^L &=& \sum_{m=0}^{n-k} \alpha_m F_S^L \otimes E^{(n-k)}_m \\
&=& \sum_{m=0}^{n-k} \alpha_m F_S^L \otimes (E_0 \otimes E^{(n-k-1)}_m + E_1 \otimes E^{(n-k-1)}_{m-1}) \\
&\deltamaps& \sum_{m=0}^{n-k} \alpha_m F_S^L \otimes E_0 \otimes (E^{(n-k-1)}_m - E^{(n-k-1)}_{m-1}) \\
= (\tG_S^L)_1 &=& \sum_{m=0}^{n-k} (\alpha_m - \alpha_{m+1}) F_S^L \otimes E_0 \otimes E^{(n-k-1)}_m \enspace.
\end{eqnarray*}
Therefore $(\tG_S^L)_1$ is of the same form as $\tG_S^L$, but with coefficients $(\alpha_m - \alpha_{m+1})$ instead of $\alpha_m$ and on one dimension less. We get the required estimate from \reflem{normGamma}(b).

There are $\binom k {k-d}$ sets $L\subset [k]$ of size $|L|=k-d$.  Since $k,d = O(1)$, one more application of the triangle equality gets the claimed bound.
\pfend

\goodbreak
\subsection{Optimization of {\large $\alpha_m$}}
\label{sec:alpha}

To maximize the adversary bound, we maximize $\|\tGamma\|$ while keeping $\|\tGamma_1\|=O(1)$.  That means, we choose the coefficients $\{\alpha_m\}$ to maximize $\alpha_0 n^{k/2}$ (\reflem{normGamma}) so that, for every $m$, $\alpha_m \le {m^{-d/2} n^{(d+1-k)/2}}$ and $\alpha_m \le \alpha_{m+1} + {n^{-k/2}}$ (\reflem{normGamma1}).


For every $r \in [n]$, $\alpha_0 \le \alpha_r + r {n^{-k/2}} \le {r^{-d/2} n^{(d+1-k)/2}} + r {n^{-k/2}}$.
The expression on the right-hand side achieves its minimum, up to a constant, $\alpha_0 = 2\ n^{(d+1)/(d+2) - k/2}$ for $r=n^{(d+1)/(d+2)}$.  This corresponds to the following solution:
\begin{equation}
\label{eqn:alphas}
\alpha_m = \max\left\{2 - \frac m {n^{(d+1)/(d+2)}}, 0\right\} n^{(d+1)/(d+2) - k/2}
\end{equation}
With this choice of $\alpha_m$, $\|\tGamma\| = \Omega(\alpha_0 n^{k/2}) = \Omega(n^{(d+1)/ (d+2)})$.

\subsection{Constructing {\large $\Gamma$} from {\large $\tGamma$}}
\label{sec:Gamma}

The matrix $\tGamma$ gives us the desired ratio of norms of $\tGamma$ and $\tGamma\circ\Delta_i$. Unfortunately, $\tGamma$ cannot directly be used as an adversary matrix, because it contains invalid columns $y$ with $f(y)=1$, that is, $y$ that contain an element of the orthogonal array on $S \subset [n]: |S|=k$, i.e., $y_S \in T_S$. We show that after removing the invalid columns the adversary matrix $\Gamma$ is still good enough.

\begin{lemma}
Let $\Gamma$ be the sub-matrix of $\tGamma$ with the invalid columns removed.  Then $\|\Gamma \circ \Delta_1\| \le \|\tGamma \circ \Delta_1\|$, and $\|\Gamma\|$ is still $\Omega(\alpha_0 n^{k/2})$ when $q \ge n^{k/(k-d)}$.
\end{lemma}

\pfstart
We estimate $\|\Gamma\|$ from below by $w^* \Gamma w'$ using unit vectors $w, w'$ with all elements equal.  Recall Equation \refeqn{GT2}:
\[
\tG_S = \alpha_0 e_0^{\otimes(n+d-k)}(e_0^{\otimes n})^* + {\sum_{u,v}} \alpha_{|v|} (u^{L_u}\otimes v)(u\otimes v)^* \enspace,
\]
where the summation is over all $u$ and $v$ such that at least one of them contains an element different from $e_0$. The sum of each column in each of  $(u^{L_u}\otimes v)(u\otimes v)^*$ is still zero because at least one of $u^{L_u}$ or $v$ sums up to zero.  Therefore the contribution of the sum is zero regardless of which columns have been removed.

By summing over all $\binom n k$ choices of $S$, we get
\[
\|\Gamma\| \ge w^* \Gamma w' = \sqrt{\binom n k} \alpha_0\ (e_0^{\otimes n})_V^* w' \enspace,
\]
where $e_V$ denotes the sub-vector of $e$ restricted to $V$, and $V$ is the set of valid columns.  Since both $e_0$ and $w'$ are unit vectors with all elements equal, and $w'$ is supported on $V$, $(e_0^{\otimes n})_V^* w' = \sqrt{|V|/q^n}$.

Let us estimate the fraction of valid columns.  The probability that a uniformly random input $y \in [q]^n$ contains an orthogonal array at any given $k$-tuple $S$ is $q^{d-k}$.  By the union bound, the probability that there exists such $S$ is at most $\binom n k q^{d-k}$.  Therefore the probability that a random column is valid is $|V| / q^n \ge 1 - \binom n k q^{d-k}$, which is $\Omega(1)$ when $q \ge n^{k/(k-d)}$.
\pfend

Thus, with the choice of $\alpha_m$ from~\refeqn{alphas}, we have $\Adv(f) = \Omega(\alpha_0 n^{k/2}) = \Omega(n^{(d+1)/(d+2)})$. This finishes the proof of \refthm{orthogonal}.

\section{Open problems}

\begin{itemize}
\item
Our lower bound $\Omega(n^{(d+1)/(d+2)})$ for the $d$-$(X, k, 1)$ orthogonal array problem is only known to be optimal when the strength $d=k-1$.  This variant corresponds to the $k$-sum problem~\cite{bs:k-sum}, for which one can prove a matching $O(n^{k/(k+1)})$ upper bound by quantum search on the Johnson graph~\cite{ambainis:eldist}.  For the $k$-distinctness problem, which lies at the other end of the spectrum with the strength $d=1$, there is a quantum algorithm running in $O(n^{1-2^{k-2}/(2^k-1)}) = o(n^{3/4})$ queries~\cite{belovs:learningKDist}, which is polynomially faster for $k \ge 3$.  Can one close the gap, say, in the simplest case $d=1$ and $k=3$, whose complexity lies between $\Omega(n^{2/3})$ and $O(n^{5/7})$?

Our lower bound only depends on $d$ but not on $k$, as long as $k = O(1)$.  This seems unlikely to be optimal.  Can one strengthen the lower bound for larger $k$?

\item
Consider the $k$-pattern problem, i.e., the $0$-$(X, k, 1)$ orthogonal array problem.  If the patterns are consistent, then the problem is equivalent to $k$ repeated unordered searches without replacement, and its complexity is $\Theta(\sqrt n)$.  If the patterns are inconsistent, then our lower bound stays $\Omega(\sqrt n)$, but the best known upper bound is just $O(n^{k/(k+1)})$.

The inconsistent $k$-pattern problem includes several interesting problems as special cases.  For example, graph collision~\cite{mss:triangle} is a 2-pattern problem and finding an $\ell$-clique is an $\binom \ell 2$-pattern problem~\cite{belovs:personal-communication-2013}.  Given a fixed graph $(V, E)$ on $n$ vertices and an $n$-bit input $x$, the \emph{graph collision problem} is to decide whether there exists an edge $\{i, j\} \in E$ with $x_i = x_j = 1$.  Given a fixed vertex set $V$, and edges $E$ specified by an input black-box, the \emph{$\ell$-clique problem} is to decide whether the graph $(V, E)$ contains a clique of size $\ell$.  Both these problems look solely for input variables labeled by 1, and the hardness of the problem comes from the fact that not every subset of input variables is admissible.  The patterns specified for non-edges resp.~non-cliques of the graphs are labeled by a dummy symbol that is not a part of the input alphabet.

Our lower bound works regardless of whether the orthogonal arrays are consistent or not, which means that it might not be strong enough for inconsistent orthogonal arrays.  Can one prove an $\omega(\sqrt n)$ lower bound for the inconsistent $k$-pattern problem?  Proving this would be a good step towards proving an $\omega(\sqrt n)$ lower bound for graph collision.

It is conceivable that the query complexity of the $k$-pattern problem can be anything between $\Omega(\sqrt n)$ and $O(n^{k/(k+1)})$, depending on the combinatorial structure of the collection of patterns.  For a consistent collection, we get $\Theta(\sqrt n)$, and the more ``inconsistent'' the orthogonal arrays are the larger the lower bound might be.  Can one lower-bound the query complexity of the inconsistent $k$-pattern problem in terms of some positive semidefinite program simpler than the full negative-weight adversary bound?  Using duality of semidefinite programming, can one then find a matching quantum algorithm, like in Ref.~\cite{Reichardt10advtight}?

\item
It is conceivable that the learning graph for $k$-distinctness~\cite{belovs:learningKDist} can be ``interpolated'' with the learning graph for the $k$-sum problem, and solve the consistent $d$-$(X, k, 1)$ orthogonal array problem.  (Essentially, one would load the first $d$ elements normally, and the remaining $k-d$ elements with only partial uncovering of loaded elements.)  Unfortunately, there are many subtle details in the analysis of the learning graph for $k$-distinctness, which makes the task of generalizing it difficult.  If one addresses all issues, what would the complexity of the learning graph for the consistent $d$-$(X, k, 1)$ orthogonal array problem be, as a function of $d$?  It will probably not match our lower bound, since there is currently a gap even for $k$-distinctness (with $d=1$), but can one at least design a quantum algorithm that for a fixed $d>1$ runs faster than $n^{1-\Omega(1)}$ for all $k$, i.e., whose complexity doesn't approach $\Omega(n^{1-o(1)})$ when $k$ grows?

The $o(n^{3/4})$-complexity learning graph for $k$-distinctness~\cite{belovs:learningKDist} can be cast as a learning graph for the consistent 1-$(X, k, 1)$ orthogonal array problem.  The learning graph crucially depends on the consistency of the orthogonal sets.  Can one generalize this learning graph to not require consistent orthogonal sets?  This is likely to be hard, witnessed by the rich combinatorial structure of the inconsistent $k$-pattern problem.

\item
Belovs and Rosmanis have recently generalized the $\Omega(n^{k/(k+1)})$ lower bound for the $k$-sum problem~\cite{bs:k-sum} to a more general framework of certificate structures~\cite{br:power-learning-graphs}.  Roughly speaking, they show strong lower bounds for the learning graph complexity of several common \emph{certificate structures} (for example, $\tilde\Omega(n^{9/7})$ for triangle finding) and then they show that for each certificate structure there exists a black-box function with that certificate structure whose query complexity satisfies the same lower bound.  Their functions are based on orthogonal arrays of strength $k-1$ when the 1-certificate size is $k$.  Their collections of orthogonal arrays are consistent, because any collection of $(k-1)$-$(X, k, 1)$ orthogonal sets is necessarily consistent.  In the case of triangle finding, their method gives a nearly tight lower bound for the \emph{triangle sum problem}.  Can their method be combined with our result to obtain nontrivial quantum query lower bounds for functions based on orthogonal arrays of smaller strengths?

\item
Our technique relies crucially on the $n^{k/(k-d)}$ lower bound on the alphabet size.  Can one relax this bound?  This will probably require an entirely new design of the adversary matrix.

\item
We have only proved a lower bound for the $d$-$(X, k, \lambda)$ orthogonal array problem with index $\lambda = 1$.  Extending our proof to larger $\lambda$ seems straightforward.  Is there a natural problem with $\lambda>1$ for which one can prove a nontrivial lower bound?
\end{itemize}

\subsection*{Acknowledgments}

We thank Aleksandrs Belovs and Ansis Rosmanis for valuable discussions.

Most of our proofs are very similar to the corresponding proofs for the quantum query lower bound of the $k$-sum problem~\cite{bs:k-sum}.  We thank Aleksandrs Belovs for agreeing to use their proofs as the basis of our paper.

\bibliographystyle{alpha}
\bibliography{../quantum}

\end{document}